ORIGINAL PAPER

# Generalized logistic growth modeling of the COVID-19 outbreak: comparing the dynamics in the 29 provinces in China and in the rest of the world

Ke Wu · Didier Darcet · Qian Wang · Didier Sornette

**Abstract** Started in Wuhan, China, the COVID-19 has been spreading all over the world. We calibrate the logistic growth model, the generalized logistic growth model, the generalized Richards model and the generalized growth model to the reported number of infected cases for the whole of China, 29 provinces in China, and 33 countries and regions that have been or are undergoing major outbreaks. We dissect the development of the epidemics in China and the impact of the drastic control measures both at the aggregate level and within each province. We quantitatively document four phases of the outbreak in China with a detailed analysis on the heterogeneous situations across provinces. The extreme containment measures implemented by China were very effective with some instructive variations across provinces. Borrowing from the experience of China, we made scenario projections on the development of the outbreak in other countries. We identified that outbreaks in 14 countries (mostly in western Europe) have ended, while resurgences of cases have been identified in several among them. The modeling results clearly show longer after-peak trajectories in western countries, in contrast to most provinces in China where the after-peak trajectory is characterized by a much faster decay. We identified three groups of countries in different level of outbreak progress, and provide informative implications for the current global pandemic.

**Keywords** Novel coronavirus (COVID-19) · Logistic growing · Epidemic modeling · Prediction

K. Wu · Q. Wang · D. Sornette (✉)
Institute of Risk Analysis, Prediction and Management (Risks-X), Academy for Advanced Interdisciplinary Studies, Southern University of Science and Technology (SUSTech), Shenzhen, China
e-mail: dsornette@ethz.ch

K. Wu · Q. Wang · D. Sornette
Department of Management, Technology and Economics (D-MTEC), Chair of Entrepreneurial Risks, ETH Zurich, Zurich, Switzerland

D. Darcet
Gavekal Intelligence Software, Nice, France

Q. Wang
Department of Banking and Finance, University of Zurich, Zurich, Switzerland

## 1 Introduction

Starting from Hubei province in China, the novel coronavirus (SARS-CoV-2) has been spreading all over the world after 2 months of outbreak in China. Facing uncertainty and irresolution in December 2019 and the first half of January 2020, China then responded efficiently and massively to this new disease outbreak by implementing unprecedented containment measures to the whole country, including lockdown of the whole province of Hubei and putting most of other provinces in de facto quarantine mode. Since March 2020, one and a half month after the national battle against the COVID-19 epidemic, China has managed to contain the virus transmission within the country, with new daily confirmed cases in mainland China excluding Hubei in the single digit range, and with just double digit numbers in Hubei. In contrast, many other countries have had fast increasing numbers of confirmed cases since March 2020, which leads to a resurgence in China due to the imported cases from overseas. On March 11, the World Health Organization (WHO) declared the coronavirus outbreak as a global pandemic. As of July 24, there are more than 15.5 million cases confirmed in more than 210 countries and territories, with 5.5 million active cases and approximately 640 thousand deaths.

For an epidemic to develop, three key ingredients are necessary: (1) source: pathogens and their reservoirs; (2) susceptible persons with a way for the virus to enter the body; (3) transmission: a path or mechanism by which viruses moved to other susceptible persons. Numerous mechanistic models based on the classical SIR model and its extensions have been utilized to study the COVID-19 epidemic. Within such a multi-agent framework, one can detail different attributes among countries, including the demographics, climate, population density, healthcare systems, government interference, etc., which will affect the three key ingredients of the epidemic mentioned above. There is a large amount of the literature using this framework studying the past major epidemics [1–6] as well as the current COVID-19 outbreak in different regions and countries [7–11]. Notably, using such a framework, a report from Neil Ferguson at Imperial College London [12] projected future scenarios with different government strategies, which had a large impact on subsequent government policies.

The numerous underlying assumptions of this kind of models vary from one model to another, leading to a huge amount of publications presenting many types of results. These mechanistic models are useful in understanding the effect of different factors on the transmission process; however, they are highly sensitive to the assumptions on the many often subtle microscopic processes. Giving an illusion of precision, mechanistic models are sometimes quite fragile and require an in-depth understanding of the dominating processes, which are likely to be missing in the confusion of an ongoing pandemics, with often inconsistent and unreliable statistics and studies performed under strong time pressure. There is thus space for simpler and, we argue, more robust phenomenological models, which have low complexity but enjoy robustness. This is the power of coarse-graining, a well-known robust strategy to model complex system [13–15].

Sophisticated statistical models including machine learning techniques have also been utilized to study and predict the development of outbreaks in different regions. For example, Gourieroux and Jasiak used time-varying Markov process to analyze the COVID-19 data [16]; Matthew Ekum and Adeyinka Ogunsanya used hierarchical polynomial regression models to predict transmission of COVID-19 at global level [17]. These methods could potentially provide good prediction results without detailed assumptions of underlying process; however, it is sometimes difficult to interpret the results and understand the fundamental dynamics.

In this paper, we focus on using phenomenological models without detailed microscopic foundations, but which have the advantage of allowing simple calibrations to the empirical reported data and providing transparent interpretations. Phenomenological approaches for modeling disease spread are particularly suitable when significant uncertainty clouds the epidemiology of an infectious disease, including the potential contribution of multiple transmission pathways [18]. In these situations, phenomenological models provide a starting point for generating early estimates of the transmission potential and generating short-term forecasts of epidemic trajectory and predictions of the final epidemic size [18].

We employ the classical logistic growth model, the generalized logistic model (GLM), the generalized Richards model (GRM) and the generalized growth model (GGM), which have been successfully applied to describe previous epidemics [18–22]. All these models have some limitations and are only applicable in some stages of the outbreak, or when enough data points are available to allow for sufficiently stable calibration. For example, an epidemic follows an exponential or quasi-exponential growth at an early stage (following the law of proportional growth with multiplier equal to the basic reproduction number $R_0$), qualifying the generalized growth model as more suitable in the initial regime. Then, the growth rate decays as fewer susceptible people are available for infection and countermeasures are introduced to hinder the transmission of the virus, so the logistic type of models is better in this later stage.

Our analysis dissects the development of the epidemics in China and the impact of the drastic control measures both at the aggregate level and within each province. Borrowing from the experience of China, we analyze the development of the outbreak in other countries. Our study employs simple models to quantitatively document the effects of the Chinese containment measures against the SARS-CoV-2 virus, and provide informative implications for the current global pandemic. The quantitative analysis allows estimating the progress of the outbreak in different countries, differentiating between those that are at a quite advanced stage and close to the end of the epidemics from those that are still in the middle of it. This is useful for scientific colleagues and decision makers to understand the different dynamics and status of countries and regions in a simple way.

The paper is structured as follows: In Sect. 2 and 3, we explain the data and the models in detail. In Sect. 4 and 5, we calibrate different models to the reported number of infected cases in the COVID-19 epidemics from January 19 to March 10 for the whole of China and 29 provinces in mainland China. Then, in Sect. 6, we perform a similar modeling exercise on other countries that have been or are undergoing major outbreaks of this virus. We discuss several limitations of our methods in Sect. 7 and then conclude in Sect. 8.

## 2 Data

### 2.1 Confirmed cases

We focus on the daily data of confirmed cases. For data from mainland China, the data source is national and provincial heath commission. We exclude the epicenter province, Hubei, which had a significant issue of underreporting at the early stage and also data inconsistency during mid-February due to a change of classification guidelines. For the provinces other than Hubei, the data are consistent except for one special event on February 20 concerning the data coming from several prisons.

We do not include the Chinese domestic data after March 10 because we conclude that the major outbreak between January and March was contained and finished. Although there have been resurgences of cases after mid-March due to imported cases from overseas countries, it is another transmission dynamics compared with the January–March major outbreak, and the risk of another round of epidemic is low given the continuing containment measures and massive testing programs [23].

For data in other countries, the source is the European Centre for Disease Prevention and Control (ECDC) [24], which is updated every day at 1 pm CET, reflecting data collected up to 6:00 and 10:00 CET. Note that the cases of the Diamond Princess cruise are excluded from Japan, following the WHO standard.

### 2.2 Data adjustment

On February 20, for the first time, infected cases in the Chinese prison system were reported, including 271 cases from Hubei, 207 cases from Shandong, 34 cases from Zhejiang. These cases were concealed before this announcement because the prison system was not within the coverage of each provincial health commission system. Given that the prison system is relatively independent and the cases are limited, we remove these cases in our data for the modeling analysis to ensure consistency.

### 2.3 Migration data

The population travels from Hubei and Wuhan to other provinces from January 1 to January 23 are retrieved from the Baidu Migration Map (http://qianxi.baidu.com).

## 3 Method

At an early stage of the outbreak, an exponential or generalized exponential model can be used to describe the data, which is intuitive and easy to calibrate. This has been employed to describe the initial processes of the epidemic in many cases, including influenza, Ebola, foot-and-mouth disease, HIV/AIDS, plague, measles, smallpox [20] and also COVID-19 [25]. The generalized growth model (GGM) that we consider is defined as:

$$\frac{dC(t)}{dt} = rC^p(t), \tag{1}$$

where $C(t)$ represents the cumulative number of confirmed cases at time $t$, $p \in [0,1]$ is an exponent that allows the model to capture different growth profiles including the constant incidence ($p=0$), sub-exponential growth ($0<p<1$) and exponential growth ($p=1$). In the case of exponential growth, the solution is $C(t) = C_0 e^{rt}$, where $r$ is the growth rate and $C_0$ is the initial number of confirmed cases at the time when the count starts. For $0<p<1$, the solution of Eq. (1) is $C(t) = C_0\left(1 + \frac{rt}{A}\right)^b$, where $b = \frac{1}{1-p}$ and $A = \frac{C_0^{1-p}}{1-p}$, so that $r$ controls the characteristic time scale of the dynamics. Essentially, the (quasi) exponential model provides an upper bound for future scenarios by assuming that the outbreak continues to grow following the same process as in the past.

However, an outbreak will slow down and reach its limit with decaying transmission rate in the end, resulting in the growth pattern departing from the (sub-)exponential path as the cumulative number of cases approaches its inflection point and the daily incidence curve approaches its maximum. Then, a logistic type model could have a better performance. In fact, the exponential model and the classical logistic model are the first- and second-order approximations to the growth phase of an epidemic curve produced by the standard Kermack–McKendrick SIR model [26, 27]. To account for subtle differences in the dynamics of different stages of an epidemic, we use three types of logistic models to describe the outbreak beyond the early growth stage:

- Classical Logistic growing model:

$$\frac{dC(t)}{dt} = rC(t)\left(1 - \frac{C(t)}{K}\right) \tag{2}$$

- Generalized Logistic model (GLM):

$$\frac{dC(t)}{dt} = rC^p(t)\left(1 - \frac{C(t)}{K}\right) \tag{3}$$

- Generalized Richards model (GRM):

$$\frac{dC(t)}{dt} = r[C(t)]^p\left(1 - \left(\frac{C(t)}{K}\right)^\alpha\right) \tag{4}$$

These three models all include two common parameters: a generalized growth rate $r$ setting the typical time scale of the epidemic growth process and the final capacity $K$, which is the asymptotic total number of infections over the whole epidemics. In the generalized logistic model, one additional parameter $p \in [0,1]$ is introduced on top of the classical logistic model to capture different growth profiles, similar to the generalized growth model (1). In the generalized Richards model, the exponent $\alpha$ is introduced to measure the deviation from the symmetric S-shaped dynamics of the simple logistic curve. The GRM recovers the original Richards model [28] for $p = 1$, and reduces to the classical logistic model (2) for $\alpha = 1$ and $p = 1$. Therefore, the GRM is more pertinent when calibrating data from a region that has entered the after-peak stage, to better describe the after-peak trajectory that may have deviated from the classical logistic decay due, for instance, to various containment measures. However, this more flexible model leads to more unstable calibrations if used on early stage data.

For the calibrations performed here, we use the standard Levenberg–Marquardt algorithm to solve the nonlinear least square optimization for the incidence curve. This is achieved by searching for the set of parameters $\hat{\Theta} = \left(\hat{\theta}_1, \hat{\theta}_2, \ldots, \hat{\theta}_m\right)$ that minimizes the sum of squared errors $\sum_{t=1}^{T}(c_t - f(t, \Theta))^2$, where $f(t, \Theta)$ is the model solution and $c_t$ is the observed data. For the fitting of the classical logistic growth function, we free the initial point $C_0$ and allow it to be one of the 3 parameters $\Theta_{\log i} = (C_0, K, r)$ to be calibrated, as the early stage growth does not follow a logistic growth. However, for the fitting of the remaining three models, $C_0$ is fixed at the empirical value. To estimate the uncertainty of our model estimates, we use a bootstrap approach with a negative binomial error structure $NB(\mu_t, \sigma_t^2)$, where $\mu_t$ and $\sigma_t^2$ are the mean and variance of the distribution at time $t$, estimated from the empirical data. Concretely, we employ the following bootstrap step to simulate $S = 500$ time series:

*Step 1* Search for the set of parameters $\hat{\Theta} = \left(\hat{\theta}_1, \hat{\theta}_2, \ldots, \hat{\theta}_m\right)$ that minimizes the sum of squared errors $\sum_{t=1}^{T} (c_t - f(t, \Theta))^2$.

*Step 2* Each simulated time series $f_i^*(t, \hat{\Theta}), i = 1, 2, \ldots, S$ is generated by assuming a negative binomial error structure as $NB(\mu_t, \sigma_t^2)$, where $\mu_t = f(t, \hat{\Theta})$ and $\sigma_t^2 = \frac{f(t, \hat{\Theta})}{T} \sum_{l=1}^{T} \frac{(c_l - f(l, \hat{\Theta}))^2}{c_l}, t = 1, 2, \ldots, T$. Thus, the probability of success $p_t = 1 - \frac{\mu_t}{\sigma_t^2} = 1 - T\left(\sum_{l=1}^{T} \frac{(c_l - f(l, \hat{\Theta}))^2}{c_l}\right)^{-1}$ in a classical negative binomial distribution is the same across $t$.

*Step 3* For each simulated time series, the parameter set $\hat{\Theta}_i, i = 1, 2, \ldots, S$ is estimated as in Step 1. Thus, the empirical distribution, correlations and confidence intervals of the parameters and the model solution can be extracted from $\hat{\Theta}_i$ and $f_i^*(t, \hat{\Theta}_i)$, where $i = 1, 2, \ldots, S$.

In the next section, we apply the most flexible model, the generalized Richards model (GRM), to study the 29 provinces in China where the outbreak is at the end. The GRM has four free parameters ($\Theta_{GRM} = (K, r, p, \alpha)$) and is able to characterize the different epidemic patterns developed in the 29 provinces. We also fit the classical logistic model to the daily incidence data as a comparison with the GRM, and a simple exponential decay model to the growth rate of cumulative confirmed cases to provide another perspective.

In Sect. 6, we apply the four models (Eqs. 1–4) to various countries and regions to identify their epidemic progress and potential future scenarios. Logistic type models tend to underestimate the final capacity $K$ and thus could serve as lower bounds of the future scenarios [29, 30]. The classical logistic model is the least flexible one among the three and usually provides the lowest estimate of the final capacity, because it fails to account for (1) the sup-exponential growth which could be captured by the GLM; (2) the potential slow abating of the epidemic which could be captured by the GRM. Both factors will increase the estimated final total confirmed numbers and they both require more data to calibrate. The performance of more flexible models increases as more data (especially data after the inflection point of the cumulative number) become available for calibration. Given the above, we define three scenarios that can be described by these four models. The *positive scenario* is defined by the model with the second lowest predicted final total confirmed cases $K$ among the three Logistic models, and the *medium scenario* is described by the model with the highest predicted final total deaths among the three Logistic models. It is important to note that both *positive* and *medium scenarios* could underestimate largely the final capacity, especially at the early stage of the epidemics. The *negative scenario* is described by the generalized growth model, which should only describe the early stage of the epidemic outbreak and is therefore least reliable for countries in the more mature stage as it does not include a finite population capacity.

## 4 Analysis at the global and provincial level for China (excluding Hubei)

### 4.1 Analysis at the aggregate level of mainland China (excluding Hubei)

As of March 10, 2020, there were in total 13,172 infected cases reported in the 30 provinces in mainland China outside Hubei. The initially impressive rising statistics have given place to a tapering associated with the limited capacity for transmission, exogenous control measure, and so on. In Fig. 1, the trajectory of the total confirmed cases, the daily increase of confirmed cases and the daily growth rate of confirmed cases in whole China excluding Hubei province are presented. The fits with the generalized Richards model and with the classical logistic growth model are shown in red and blue lines, respectively, in the upper, middle and lower left panel, with the data up to March 1, 2020. In the lower left panel of Fig. 1, the daily empirical growth rate $r(t) = \log \frac{C(t)}{C(t-1)}$ of the confirmed cases is plotted in log scale against time. We can observe two exponential decay regimes of the growth rate with two different decay parameters before and after February 14, 2020. The green line is the fitted linear regression line (of the logarithm of the growth rate as a function

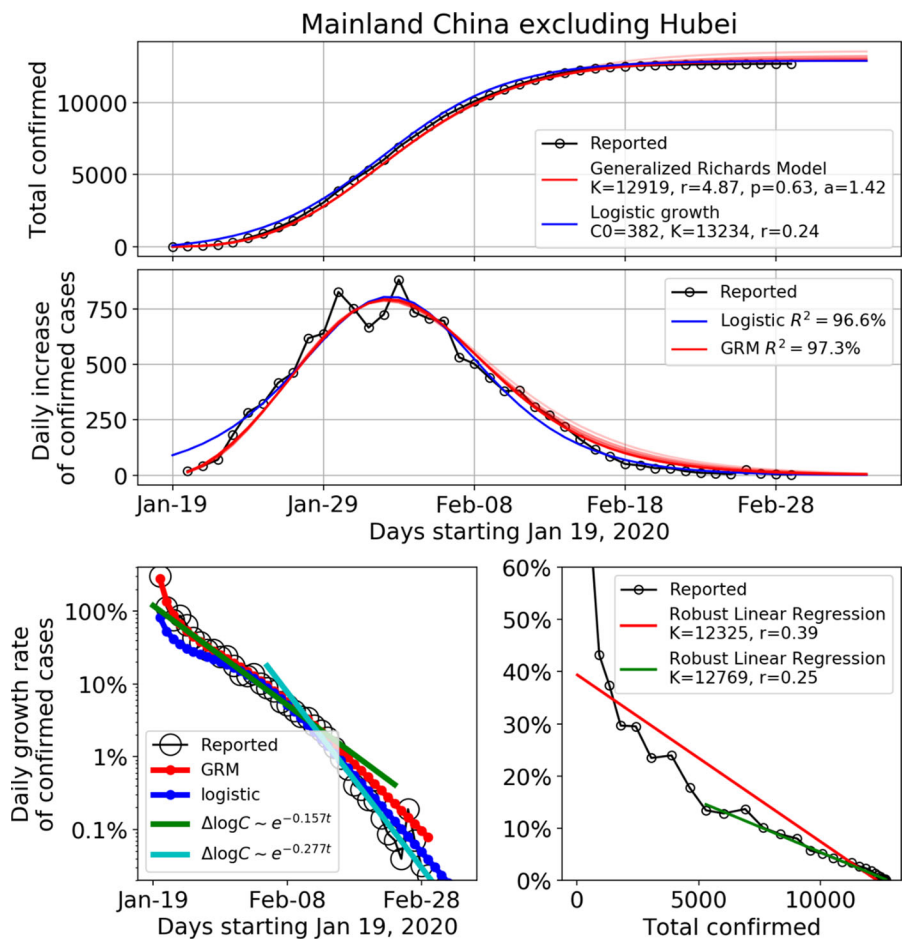

**Fig. 1** Time dependence of the total number of confirmed cases (upper panel), the daily number of new confirmed cases (middle panel), and the daily growth rate of confirmed cases (lower panel) in the mainland China excluding Hubei province until March 1, 2020. The empirical data are marked by the empty circles. The blue and red lines in the upper, middle and lower left panels show the fits with the logistic growth model and generalized Richards model (GRM) respectively. For the GRM, we also show the fits using data ending 20, 15, 10, 5 days earlier than March 1, 2020, as lighter red lines in the upper and middle panel. This demonstrates the consistency and robustness of the fits. The lower left panel shows the daily growth rate of the confirmed cases in log scale against time. The green and cyan straight lines show the linear regression of the logarithm of the growth rate as a function of time for the period of January 25 to February 14, and the period of February 15 to March 1, respectively. The lower right panel is the daily growth rate of the confirmed cases in linear scale against the cumulative number of confirmed cases. The red and green lines are the linear fits for the period of January 19 to February 1, and the period of February 2 to March 1, respectively

of time) for the data from January 25 to February 14, 2020, yielding an exponential decay parameter equal to −0.157 per day (95% CI: (−0.164, −0.150)). This indicates that after the lockdown of Wuhan city on January 23 and the top-level health emergency activated in most provinces on January 25, the transmission in provinces outside Hubei has been contained with a relatively fast exponential decay of the growth rate from a value starting at more than 100% to around 2% on February 14. Then, starting February 15, 3 weeks after a series of extreme controlling measures, the growth rate is found to decay with a faster rate with a decay parameter equal to −0.277 per day (95% CI: (−0.313, −0.241)).

This second regime is plotted as the cyan line in the lower left panel of Fig. 1. The green and cyan straight lines show the linear regression of the logarithm of the growth rate as a function of time for the period of January 25 to February 14, and the period of February 15 to March 1, respectively. The asymptotic exponential decay of the growth rate can be justified theoretically from the generalized Richards model (4) by expanding it in the neighborhood where $C$ converges to $K$. Introducing the change of variable $C(t) = K(1 - \varepsilon(t))$, and keeping all terms up to first order in $\varepsilon(t)$, Eq. (4) yields

$$\frac{d\varepsilon(t)}{dt} = -\gamma\varepsilon(t) \quad \text{with} \quad \gamma = r\alpha K^{p-1}. \quad (5)$$

This gives

$$\frac{1}{C}\frac{dC(t)}{dt} = \frac{\epsilon_0 \gamma e^{-\gamma t}}{1 - \epsilon_0 e^{-\gamma t}} = \gamma\left(\epsilon_0 e^{-\gamma t} + [\epsilon_0 e^{-\gamma t}]^2 + [\epsilon_0 e^{-\gamma t}]^3 + \cdots\right), \quad (6)$$

where $\varepsilon_0$ is a constant of integration determined from matching this asymptotic solution with the non-asymptotic dynamics far from the asymptote. Thus, the leading behavior of the growth rate at long times is $\frac{1}{C}\frac{dC(t)}{dt} = \gamma\epsilon_0 e^{-\gamma t}$, which is exponential decaying as shown in the lower left panel of Fig. 1. Using expression (5) for $\gamma$ as a function of the 4 parameters $r, \alpha, K$ and $p$ given in the inset of the top panel of Fig. 1, we get $\gamma = 0.21$ for mainland China excluding Hubei, which is bracketed by the two fitted values 0.17 and 0.28 of the exponential decay given in the inset of the lower left panel of Fig. 1.

In the lower right panel of Fig. 1, the empirical growth rate $r(t)$ is plotted in linear scale against the cumulative number of confirmed cases. The red and green lines are the linear regressed lines for the full period and for the period after February 1, 2020, respectively. We can see that the standard logistic growth cannot capture the full trajectory until February 1. After February 1, the linear fit is good, qualifying the simple logistic equation ($p=1$ and $\alpha=1$), with growth rate $r$ estimated as 0.25 for the slope, which is compatible with the value determined from the calibration over the full data set shown in the top two panels of Fig. 1.

Figure 2a demonstrates the sensitivity of the calibration of the GRM to the end date of the data by presenting six sets of results for six end dates. Specifically, the data on the daily number of new confirmed case are assumed to be available until 23 January, 28 January, 2 February, 7 February, 12 February, 17 February, i.e., 30, 25, 20, 15, 10 and 5 days before February 22 were presented. For each of the six data sets, we generated 500 simulations of $\frac{dC(t)}{dt}$ based on the best fit parameters using parametric bootstrap with a negative binomial error structure, as in prior studies [20]. Each of these 500 simulations constitutes a plausible scenario for the daily number of new confirmed cases, which is compatible with the data and GRM. The dispersion among these 500 scenarios provides a measure of stability of the fits, and their range of values gives an estimation of the confidence intervals. The first conclusion is the non-surprising large range of scenarios obtained when using data before the inflection point, which, however, encompass the realized data. We observe a tendency for early scenarios to predict a much faster and larger number of new cases than observed, which could be expected in the absence of strong containment control. With more data, the scenarios become more accurate, especially when using realized data after the peak, and probably account well for the impact of the containment measures that modified the dynamics of the epidemic spreading. This is confirmed again in Fig. 2b, which presents the convergences of the four parameters of the GRM. The confidence intervals decrease significantly once the data are available after the inflection point.

### 4.2 Analysis at the provincial level (29 provinces) of mainland China (excluding Hubei)

As of March 1, 2020, the daily increase of the number of confirmed cases in China excluding Hubei province has decreased to less than 10 cases per day. The preceding one-month extreme quarantine measures thus seem to have been very effective from an aggregate perspective, although there is a resurgence of cases since mid-March due to imported cases from overseas countries. At this time, it is worthwhile to take a closer look at the provincial level to study the effectiveness of measures in each province. The supplementary material presents figures similar to Fig. 1 for each of the 29 provinces in mainland China, and a table presenting some useful statistics for each province and the values of the fitted parameters of the generalized Richards model, logistic growth model and the exponential decay exponent of the growth rate. Tibet is excluded as it only has 1 confirmed case as of March 10. This analysis at the 29 provinces allows us to identify four phases in the development of the epidemic outbreak in mainland China.

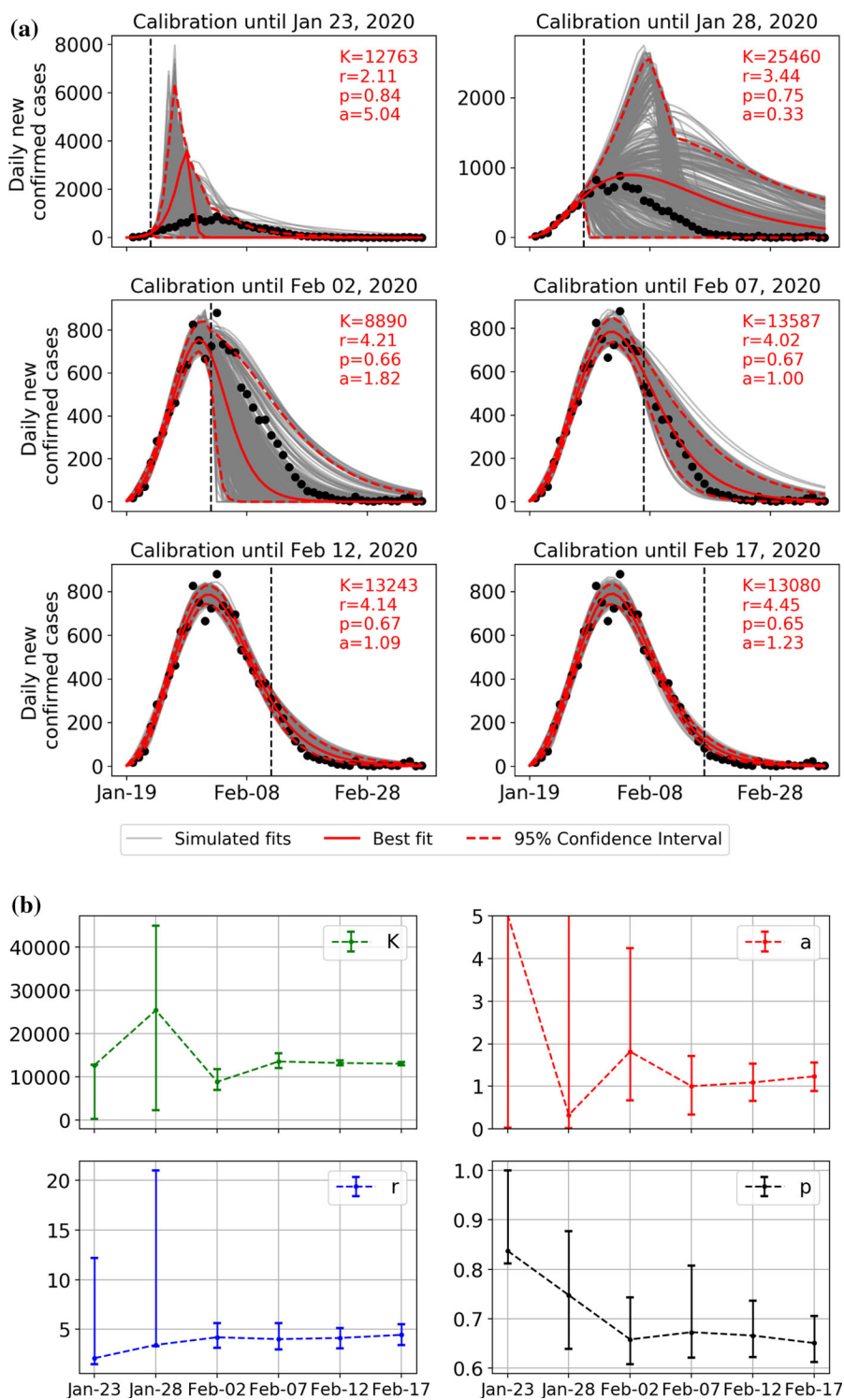

**Fig. 2 a** Daily number of new observed confirmed cases for mainland China excluding Hubei (black circles) compared with 500 scenarios built by parametric bootstrap with a negative binomial error structure on the GRM model with best fit parameters determined on the data up to the time indicated by the vertical dashed line. The last time used in the calibration is, respectively, 5, 10, 15, 20, 25, 30 days before February 22, 2020 from bottom to top. The red continuous line is the best fitted line, and the two dashed red curves delineate the 95% confidence interval extracted from the 500 scenarios. The six panels correspond each to a different end date, shown as the sub-title of each panel, at which the data have been calibrated with the GRM model. **b** Convergence of the four parameters from the GRM simulations shown in **a**. The error bars indicate the 80% prediction intervals. (Color figure online)

- *Phase I (January 19–January 24, 6 days): early stage outbreak* The data mainly reflect the situation before January 20, when no measures were implemented, or they were of limited scope. On January 19, Guangdong became the first province to declare a confirmed case outside Hubei in mainland China [22]. On January 20, with the speech of President Xi, all provinces started to react. As of January 24, 28 provinces reported confirmed cases with daily growth rates of confirmed cases ranging from 50% to more than 100%.
- *Phase II (January 25–February 1, 8 days) fast growth phase approaching the peak of the incidence curve (inflection point of the cumulative number)* The data start to reflect the measures implemented in the later days of Phase I and in Phase II. In this phase, the government measures against the outbreak have been escalated, marked by the lockdown of Wuhan on January 23, the top-level public health emergency state declared by 20+ provinces by January 25, and the standing committee meeting on January 25, the first day of the Chinese New Year, organized by President Xi, to deploy the forces for the battle against the virus outbreak. In this phase, the growth rate of the number of confirmed cases in all provinces declined from 50 to 10%+, with an exponentially decay rate of 0.157 for the aggregated data. At the provincial level, some provinces failed to see a continuous decrease in the growth rate and witnessed the incidence grow at a constant rate for a few days, implying exponential growth of the confirmed cases. These provinces include Jiangxi (~40% until January 30), Heilongjiang (~25% until February 5), Beijing (~15% until February 3), Shanghai (~20% until January 30), Yunnan (~75% until January 27), Hainan (~10% until February 5), Guizhou (~25% until February 1), Jilin (~30% until February 3). Some other provinces managed to decrease the growth rate exponentially during this period. As of February 1, 15 provinces had reached the peak of the incidence curve, indicating the effectiveness of the extreme measures, and most provinces started to be in control of the epidemics.
- *Phase III (February 2–February 14, 13 days) slow growth phase approaching the end of the outbreak* In this period, all provinces continued to implement their strict measures, striving to bring the epidemics to an end. The growth rate of the number of confirmed cases declined exponentially with similar rates as in Phase II, pushing down the growth rate from 10 to 1%. In phase III, all provinces have passed the peak of the incidence curve, which allows us to obtain precise scenarios for the dynamics of the end of the outbreak from the model fits (Fig. 2). As of February 14, 23 out of 30 provinces have less than 10 new cases per day.
- *Phase IV (February 15–8 March) the end of the outbreak* Starting February 15, the exponential decay of the growth rate at the aggregate level has switched to an even faster decay with parameter of 0.277 (Fig. 1). As of February 17, 1 week after normal work being allowed to resume in most provinces, 22 provinces have a growth rate smaller than 1%. As of February 21, 28 provinces have achieved 5-day average growth rates smaller than 1%.

## 5 Analysis of the development of the epidemic and heterogeneous Chinese provincial responses

### 5.1 Quantification of the initial reactions and ramping up of control measures

On January 19, Guangdong was the first province to report a confirmed infected patient outside Hubei. On January 20, 14 provinces reported their own first case. During January 21–23, another 14 provinces reported their first cases. If we determine the peak of the outbreak from the 5 days moving average of the incidence curve, then there are 15 provinces taking 7–11 days from their first case to their peak, 9 provinces taking 12–15 days and 6 provinces taking more than 15 days. If we define the end of the outbreak as the day when the 5 days moving average of the growth rate becomes smaller than 1%, then 7 provinces spent 8–12 days from the peak to the end, 7 provinces spent 13–16 days, 13 provinces spent 17–20 days and 2 provinces spent 21–22 days. For the six provinces that have the longest duration from the start of their outbreak to the peak (more than 15 days), it took 8–13 days for them to see the end of the outbreak (Fig. 3). This means that these 6 provinces were able

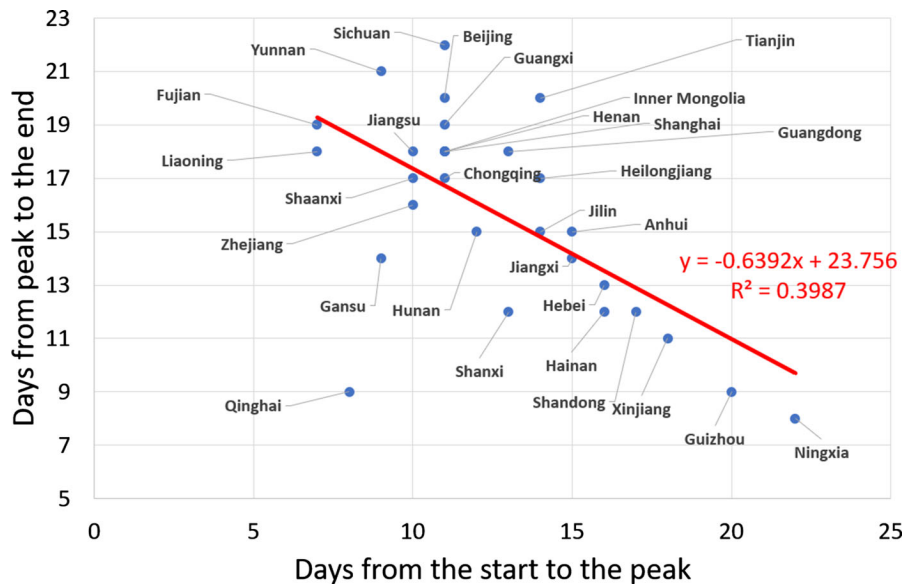

**Fig. 3** Inverse relationship found across the 29 Chinese provinces between the number of days from peak to the end and the duration from start to the peak of the epidemics. Here, the end of the outbreak is defined operationally as the day when the 5 days moving average of the growth rate becomes smaller than 1%

to control the local transmissions of the imported cases quite well, so that the secondary transmissions were limited. In contrast, 20 provinces took 28–31 days from the start to the end of the outbreak. Thus, those provinces that seem to have responded sluggishly during the early phase of the epidemics seem to have ramped up aggressively their countermeasures to achieve good results.

### 5.2 Diagnostic of the efficiency of control measures from the exponential decay of the growth rate of infected cases

The 10 most infected provinces (Guangdong, Henan, Zhejiang, Hunan, Anhui, Jiangxi, Shandong, Jiangsu, Chongqing, Sichuan) have done quite well in controlling the transmission, as indicated by the fact that their daily growth rates follow well-defined exponential decays, with all of their $R^2$ larger than 90%. This exponential decay continued for all ten provinces until the situation was completely under control during February 15–18, when the daily incidence was at near zero or a single-digit number. Eight out of these ten provinces have an exponential decay exponent of the growth rate ranging from 0.142 to 0.173, similar to what is observed at the national average level (0.157). Note that this exponential decay can be inferred from the generalized Richards model, as we noted in Eqs. (5) and (6) in Sect. 4.1.

### 5.3 Zhejiang and Henan exemplary developments

Zhejiang and Henan are the second and third most infected provinces but have the fastest decaying speed of the incidence growth rate (exponential decay exponent for Zhejiang: 0.223, Hunan: 0.186) among the most infected provinces. This is consistent with the fast and strong control measures enforced by both provincial governments, which have been praised a lot on Chinese social network [31, 32]. As one of the most active economies in China and one of the top provinces receiving travelers from Wuhan around the Lunar New Year [33], Zhejiang was the first province launching the top-level public health emergency on January 23, and implemented strong immediate measures, such as closing off all villages in some cities. The fitted curves from the GRM and logistic growth models indicate a peak of the incidence curve on January 31, which is the earliest time among top infected provinces. Similarly, Henan Province, as the neighbor province of Hubei and one of the most populated provinces in China, announced the suspension of passenger bus to and from Wuhan at the end of December 2019. In early January 2020, Henan implemented a series of actions including suspending poultry trading, setting up return spots at the village entrances for people from Hubei, listing designated hospitals for COVID-19 starting as early as January 17, and so on [32]. These actions were the first to be implemented among all provinces.

### 5.4 Heterogeneity of the development of the epidemic and responses across various provinces

Less infected provinces exhibit a larger variance in the decaying process of the growth rate. However, we also see good examples like Shanghai, Fujian and Shanxi, which were able to reduce the growth rate consistently with a low variance. These provinces benefited from experience obtained in the fight against the 2003 SARS outbreak or enjoy richer local medical resources [34]. This enabled the government to identify as many infected/suspected cases as possible in order to contain continuously the local transmissions. Bad examples include Heilongjiang, Jilin, Tianjin, Gansu, which is consistent with the analysis of [34].

Most provinces have a small parameter of $p$ in the GRM [see Eq. (4)] and an exponent α large than 1, indicating that China was successful in containing the outbreak as sub-exponential growing process ($p < 1$), with a faster than logistic decay (α > 1) in most provinces, except Guangdong, Zhejiang, Jiangxi, Sichuan, Heilongjiang, Fujian, Yunnan and Gansu. However, these exceptional provinces are due to various reasons, which may not necessarily be the ineffective measures. The large $p$ and small α in Guangdong and Zhejiang are likely due to their high population densities and highly mobile populations in mega-cities, which are factors known to largely contribute to the fast transmission of viruses. Jiangxi, Sichuan, Fujian, Yunnan and Gansu all had a fast growth phase before February 1, but were successful in controlling the subsequent development of the epidemics. The fast growth phase in Heilongjiang lasted a bit longer than the above-mentioned provinces, due to the occurrence of numerous local transmissions. Heilongjiang is the northernmost province in China, so it is far from Hubei and does not have a large number of migrating people from Hubei. However, compared with other provinces, it has a high statistic of both the confirmed cases and the case fatality rate (2.7% as of March 10), which has been criticized a lot by the Chinese social network.

### 5.5 Initial and total confirmed numbers of infected cases correlated with travel index

The initial value $C_0$ of the logistic equation could be used as an indicator of the early number of cases, reflecting the level of early contamination from Hubei province as the epicenter of the outbreak. To support this proposition, the upper panel of Fig. 4 plots the estimated $C_0$ as a function of the migration index from Hubei and Wuhan to each province. The migration index is calculated as equal to 25% of the population migrating from Hubei (excluding Wuhan) plus 75% of the population migrating from Wuhan, given that Wuhan was the epicenter and the risks from the Hubei region excluding Wuhan are lower. One can observe a clear positive correlation between the estimated $C_0$ and the migration index. The lower panel of Fig. 4 shows an even stronger correlation between the total number of cases recorded on March 6 and the travel index, expressing that a strong start of the epidemics predicts a larger number of cases, which is augmented by infections resulting from migrations out of the epidemic epicenter.

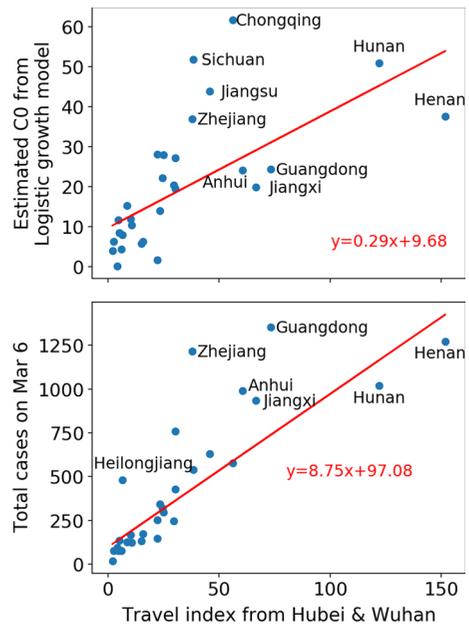

**Fig. 4** Upper panel: estimated $C_0$ for the logistic growth model versus travel index from Hubei and Wuhan. Lower panel: total confirmed cases versus travel index from Hubei and Wuhan. The Pearson correlation between $C_0$ and the migration index is 0.65 ($p < 10^{-3}$), and the correlation between the cumulative number of confirmed cases and the migration index is 0.82 ($p < 10^{-4}$)

## 6 Analysis of the epidemic in various countries

Based on the phenomenological models presented above, we have been publishing daily forecasts for the number of COVID-19 confirmed cases in various countries/regions in the world since March 23. Four months after our first daily report, the outbreaks in most European countries have come to an end and the epicenter has shifted to the USA, South Hemisphere countries and other developing countries. In this section, with the four models (Eqs. 1–4) and the resulting three scenarios we specified in Sect. 3, we first present the model outputs for countries with an ended first outbreak wave. We report that they have significantly different parameters compared to China. Then, we present the latest status of the countries in the middle of their outbreak and discuss the performance of our models.

6.1 Countries with a matured outbreak

As of July 24, Europe (including Russia) has cumulatively 3.06 million confirmed cases with a growth rate of 0.5–1% per day, gradually decreased from more than 5% 2 months ago. The outbreak in Europe started in February from Italy, France and Spain and then spread to the whole Europe. The first wave of outbreaks has ended in most West European countries, which started to ease their lockdowns in late April. However, the rising numbers of cases in the East European countries (especially Russia) and the resurgences of cases in the West European countries since June have kept the daily number of confirmed cases in Europe approximately flat after May.

As of July 24, the USA has cumulatively 4.03 million confirmed cases with a daily growth rate of around 2%, increased from 1% a month ago. The first wave of outbreaks in the USA was concentrating in New York with a national lockdown from March to late April. However, with the pressure from people in several states and from President Trump, the USA reopened since May. The ease of lockdowns and the street demonstrations associated with the BLM ("black lives matter") movement where large numbers of people gather have contributed to a second wave of outbreaks across the whole country, with the daily new cases of the USA keeping breaking records. The epicenter of this second wave of the outbreak has shifted from New York to several states such as Alabama, Arkansas, Arizona, California and Florida.

Although the flexible four-parameter generalized Richards model (GRM) can capture the slow decay of the after-peak trajectory in an epidemic, it cannot characterize the second wave dynamics by construction. Therefore, we use June 5 as the cutoff date for the data used in the exercise of model fitting to Europe as a whole and to 14 countries that have ended their first waves. These countries are the USA, 11 west European countries (the UK, Spain, Italy, Germany, France, Austria, Belgium, Netherlands, Ireland, Portugal and Switzerland) and two Asian countries (Japan and Turkey). Among these 14 countries, a second wave of outbreak has started in the USA and Japan. Resurgences of cases are identified in several countries including Spain, Germany, France, Austria, Portugal, Switzerland and Turkey.

Figure 5 presents the daily confirmed cases of Europe and of 14 countries fitted with the best two models among the four models. In contrast to the results from most provinces in China in the after-peak stage, all West European countries have a small estimated parameter α in the generalized Richards model, indicating a slow decay of the after-peak trajectory. This is possibly because the full lockdown in most West European countries was implemented when the virus transmissions have already started among local communities. The heterogeneous stage of the outbreaks and the resurgences of cases in some countries (e.g., Spain, German, France, Austria, Portugal, Switzerland) also contribute to a slow decay of Europe as a whole, which is in an approximate plateau over the past month. In contrast, the extreme lockdown and containment measures in China were implemented at the beginning of the outbreak and were maintained strictly through the whole period of the outbreak with centralized management, contributing to the fast decay of the outbreak in the after-peak stage.

The resurgences of cases and a second wave happening in the USA and various countries show that the risk of new outbreaks is not negligible. The antibody tests performed in various countries and regions [35] also show that the herd immunity level in the crowd is not yet sufficient to prevent another outbreak.

6.2 Countries in the middle of their outbreaks

In Table 1, we report the latest confirmed cases per million population and the estimated outbreak

Generalized logistic growth modeling of the COVID-19 outbreak: comparing the dynamics

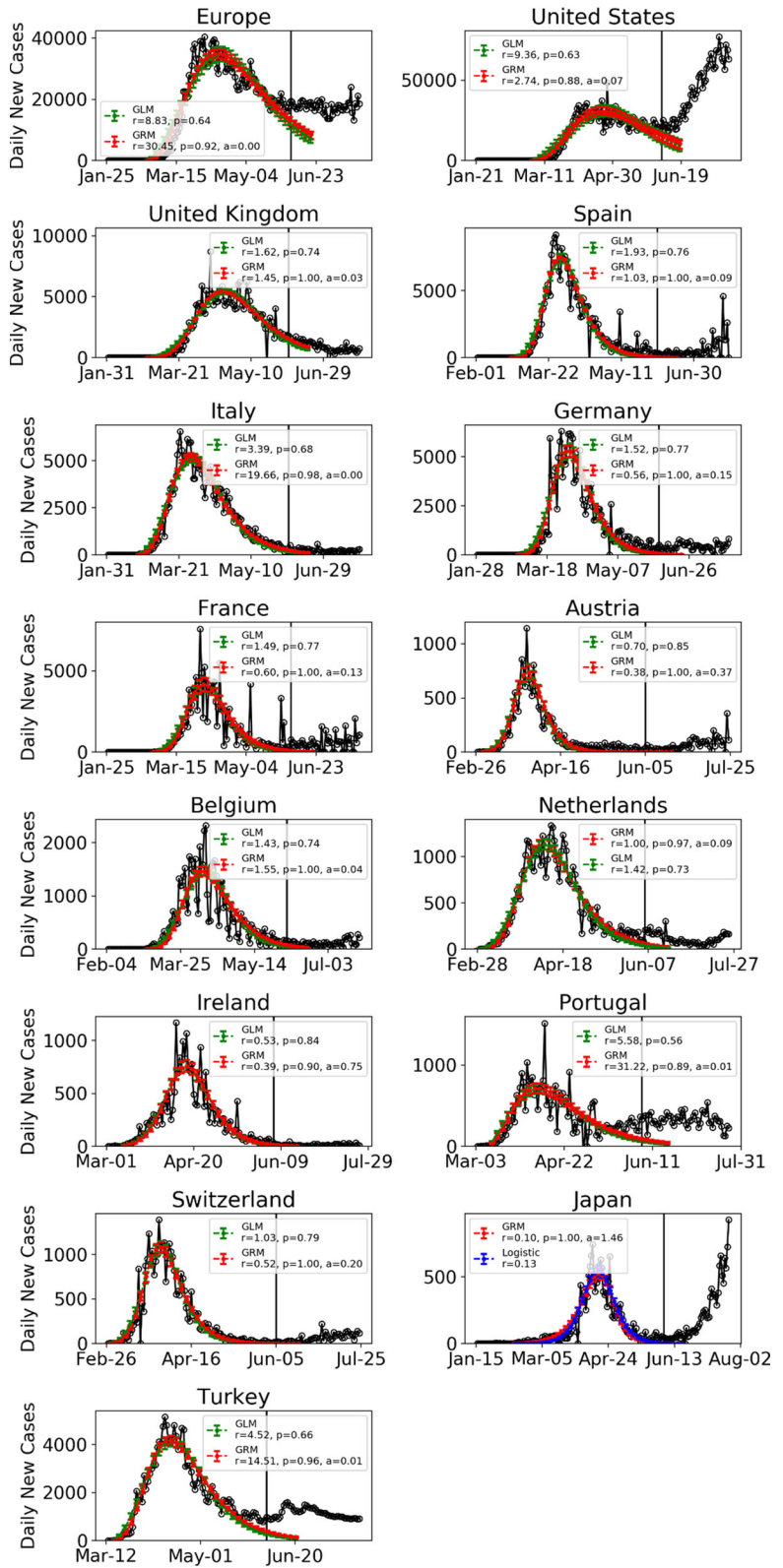

**Fig. 5** Daily confirmed cases of Europe and of 14 countries with the best two models among the four models (1)–(4). The empirical data are indicated by empty circles. The blue, red and green lines with the error bars show the fits with the logistic growth model, generalized Richards model (GRM), and generalized logistic model (GLM), respectively. The error bars indicate 80% prediction intervals. Data are plotted every 3 days. The vertical line indicates the date (June 5) up to when the data are fed to the model. (Color figure online)

**Table 1** Current confirmed cases per million population and estimated outbreak progress in positive and medium scenarios (July 24 confirmed cases divided by the estimated total final confirmed cases in positive and medium scenario)

|  | Confirmed per million population (July 24) | Outbreak progress in positive scenario | Outbreak progress in medium scenario |
|---|---|---|---|
| Chile | 18,087 | 99.4% (60.2%, 100.0%) | 98.3% (53.1%, 100.0%) |
| Canada | 3040 | 98.9% (95.2%, 100.0%) | 97.2% (93.9%, 100.0%) |
| Afghanistan | 967 | 96.9% (91.5%, 100.0%) | 96.4% (90.4%, 100.0%) |
| Qatar | 38,913 | 96.5% (94.4%, 98.4%) | 95.6% (93.3%, 97.8%) |
| Belarus | 7031 | 94.0% (92.0%, 96.3%) | 91.6% (89.5%, 93.8%) |
| Pakistan | 1274 | 95.8% (92.6%, 98.8%) | 88.2% (85.2%, 91.2%) |
| Russia | 5503 | 80.4% (75.9%, 84.7%) | 77.1% (73.6%, 80.9%) |
| Peru | 11,601 | 73.8% (68.5%, 79.1%) | 72.5% (65.7%, 79.6%) |
| Saudi Arabia | 7727 | 89.9% (85.4%, 93.7%) | 71.4% (67.3%, 75.4%) |
| Sweden | 7735 | 90.5% (82.5%, 99.8%) | 53.7% (38.7%, 65.5%) |
| Brazil | 10,920 | 54.2% (36.8%, 79.6%) | 49.6% (38.6%, 97.2%) |
| Argentina | 3189 | 30.0% (8.5%, 47.7%) | 26.9% (8.4%, 77.4%) |
| Mexico | 2938 | 35.1% (27.6%, 42.3%) | 23.9% (14.0%, 45.7%) |
| Israel | 6527 | 100.0% (23.1%, 100.0%) | Not reliable |
| India | 952 | 78.0% (17.7%, 100.0%) | 23.0% (14.3%, 31.7%) |
| Dominican Republic | 5421 | Not reliable | Not reliable |
| Philippines | 698 | Not reliable | Not reliable |
| Iran | 3472 | Not reliable | Not reliable |

The ranking is in terms of outbreak progress in medium scenario (fourth column from left). Numbers in brackets are 80% confidence intervals. As positive scenarios predict a smaller final number of total infected cases, the outbreak progress is thus larger in the positive scenario progress in the positive and medium scenarios for various countries including Chile, Canada, Afghanistan, Qatar, Belarus, Pakistan, Russia, Peru, Saudi Arabia, Brazil, Argentina, Sweden, Mexico, Israel, India, Dominican Republic, Philippines and Iran. The outbreak progress is defined as the latest confirmed cases divided by the estimated final total confirmed case, either in the positive or the medium scenario. In Fig. 6, the daily confirmed cases of these 18 countries are shown together with the best two models among the four models (1)–(4). Figure 7 presents the ensemble distribution of the estimated final total

**Fig. 6 a** Daily confirmed cases of 10 countries with the best two models among the four models (1)–(4). The empirical data are indicated by empty circles. The blue, red and green lines with the error bars show the fits with the logistic growth model, generalized Richards model (GRM) and generalized logistic model (GLM), respectively. The error bars indicate 80% prediction intervals. Data are plotted every 3 days. The vertical line indicates the date (July 24) up to when the data are fed to the model. **b** Daily confirmed cases of 8 countries with the best two models among the four models (1)–(4). The empirical data are indicated by empty circles. The blue, red and green lines with the error bars show the fits with the logistic growth model, generalized Richards model (GRM) and generalized logistic model (GLM), respectively. The error bars indicate 80% prediction intervals. Data are plotted every 3 days. The vertical line indicates the date (July 24) up to when the data are fed to the model. (Color figure online) confirmed cases per million population obtained by aggregating the positive and medium scenarios. The distributions obtained on June 6, 2020, are plotted on the right side of each violin plot in gray, while the distributions obtained on July 24 are plotted on the left side in cyan.

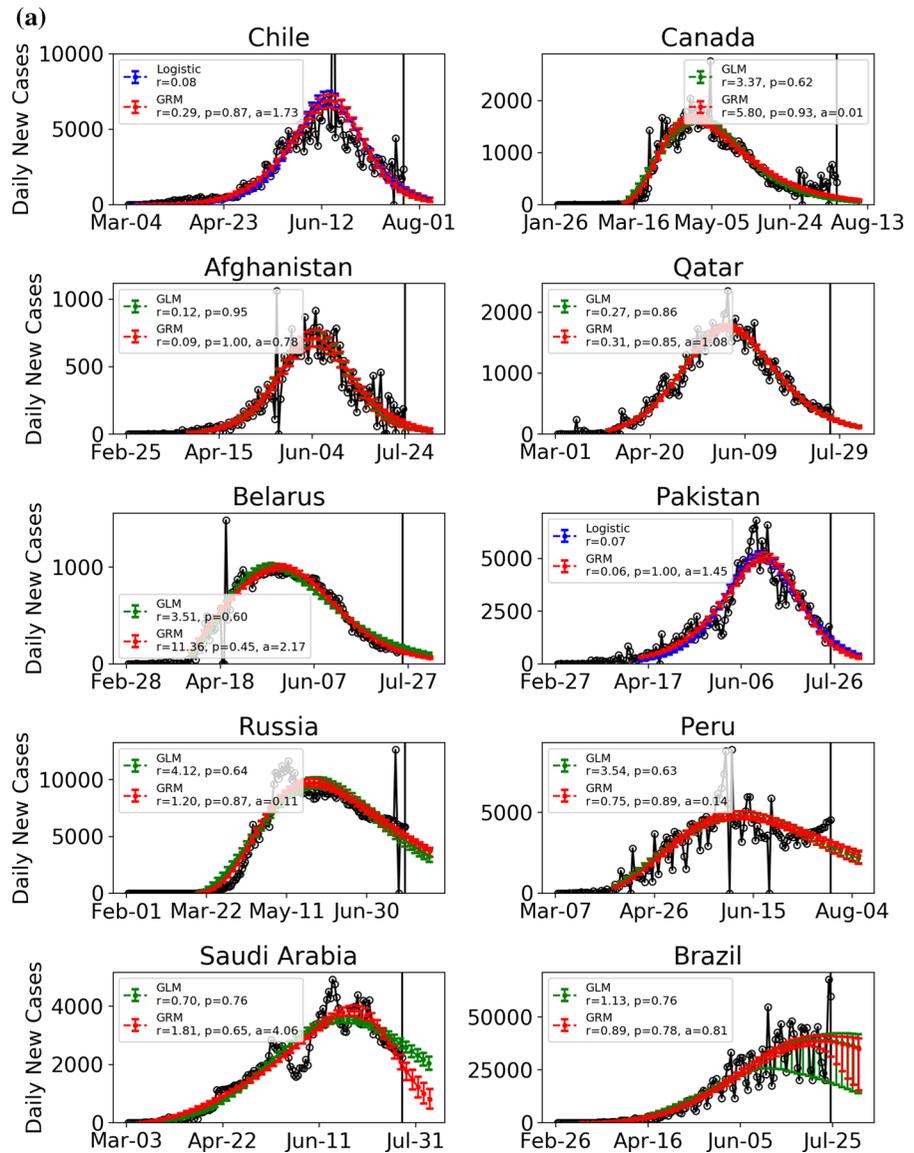

Among them, the most matured group of countries includes Chile, Canada, Afghanistan, Qatar and Belarus, which have strong signs that their inflection points have been passed with an outbreak progress larger than 90% in the medium scenario. The next group of countries includes Pakistan, Russia, Peru and Saudi Arabia, which are less matured with outbreak progress in the range 60–90% in the medium scenario. They have developed signs of passing their inflection points, but may reverse to their previous exponential growth if there is a second wave of outbreak. All these countries have their distributions of final confirmed cases converged, indicating the reliability of the future projected scenarios.

The least mature group of countries are Sweden, Brazil, Argentina, Mexico, Israel, India, Dominican Republic, Philippines and Iran. Sweden, Israel,

**Fig. 6** continued

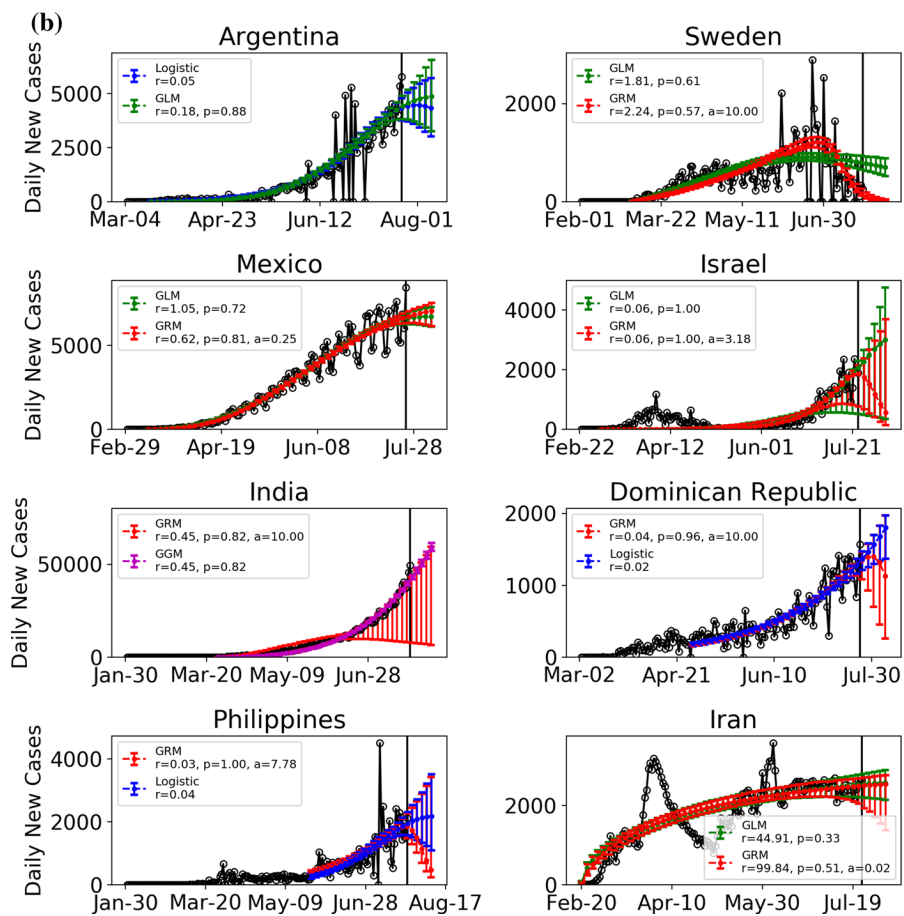

Philippines and Iran are experiencing a second outbreak, while Brazil, Argentina, Mexico, India and Dominican Republic are still in the exponential growth stage, indicating highly uncertain future projections. As shown in Fig. 7, their ensemble distributions of final confirmed cases are non-converged or highly dispersed.

As discussed before, our models are unable to characterize the second wave dynamics by construction. Thus, the results concerning the predictions for countries experiencing a second wave are not reliable. Note that the logistic type models are usually useful in understanding the short-term dynamics over a horizon of a few days, while may tend to underestimate long-term results due to the change of the fundamental dynamics resulting from government interventions, a second wave of outbreaks or other factors. This is shown by large shifts between the obtained distributions between June 5 and July 24 in Fig. 7. Thus, the estimated outbreak progress serves both as a lower bound for future developments and as an indication of what to expect of the evolution dynamics of the epidemics. To have a view of the performance of short-term predictions, we present the latest 7-day prediction errors for the total number of confirmed cases in Fig. 8, based on positive and medium scenarios. One can see that our 7-day predictions based on the data up to July 17 are correct with narrow prediction intervals in all matured countries. Our 7-day predictions, however, underestimate the realized values in immature countries including India, Argentina and Israel. Until May 24, 2020, we have uploaded a daily update of our projections and an analysis of forecasting errors online [36]. Thereafter, we shifted to a weekly update until July 3, we have now discontinued it as the epidemics have entered second waves and other regimes highly dependent on country specific characteristics.

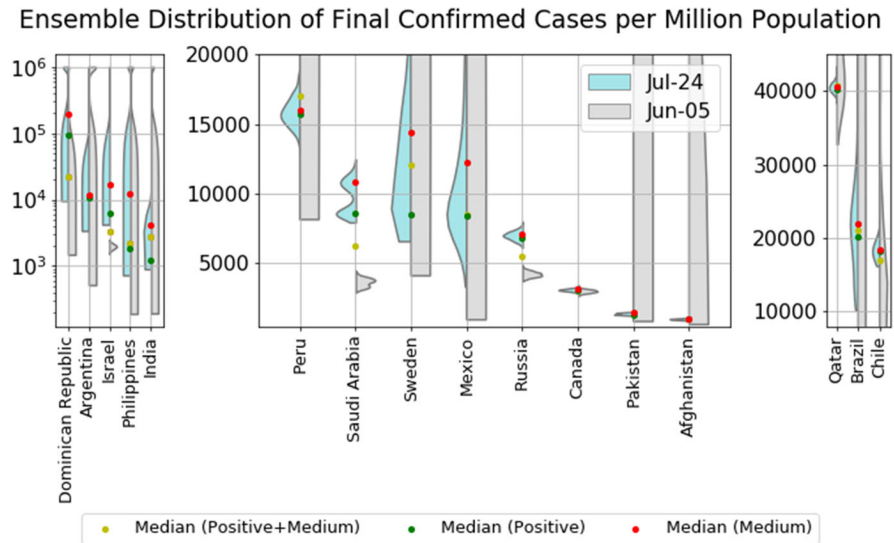

**Fig. 7** Violin plot of the distributions of the final total number of confirmed cases per million derived by combining the distributions of the positive and medium scenarios. The left side of each violin in cyan shows distributions obtained on July 24, while the right side of each violin in gray shows distributions obtained on June 5. The model setup in the negative scenario does not incorporate a maximum saturation number and thus cannot be used. The yellow dots indicate the median prediction for the combined distribution, while the green and red dots indicate the median of the positive and of the medium scenarios, respectively. (Color figure online)

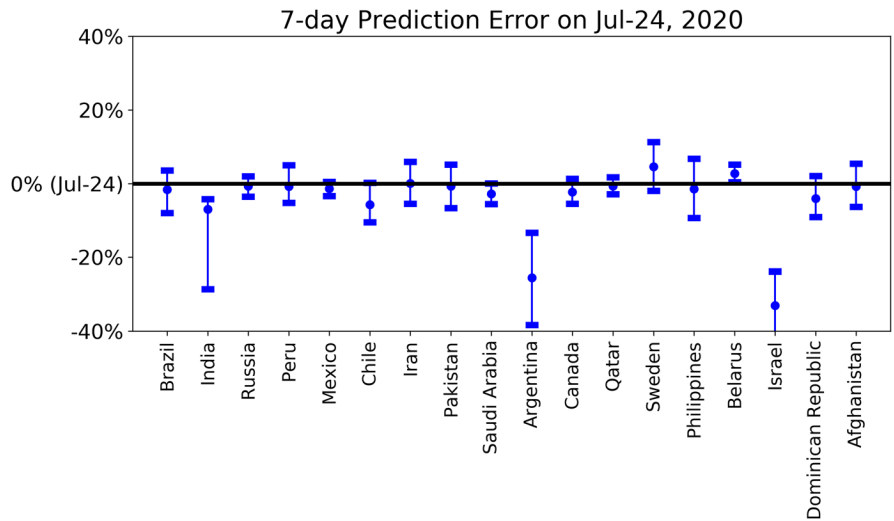

**Fig. 8** 7-day prediction error of the forecast performed on July 17 for the total number of deaths for various countries/regions. The horizontal line corresponds to the empirical data on July 24. The error bars are 80% prediction intervals, and the middle dots are the median predictions based on the predictions from the positive and medium scenarios. A negative value corresponds to a prediction that underestimated the true realized value

## 7 Discussion on the limitation of the method

In this paper, we only apply the models to the number of confirmed cases, which is largely subject to a number of extraneous variations among countries such as case definition, testing capacity, testing protocol and reporting system. It is important to note that there is a significant limitation in using this statistic to estimate the true situation of the outbreak in a country, and we lay out four major variables here:

- *Case definition* Different countries employ different definitions of a confirmed COVID-19 case, and the definition also changes over the time. For example, China's national health commission issued seven versions of a case definition for COVID-19 between 15 January and 3 March, and a recent study found each of the first four changes increased the proportion of cases detected and counted by factors between 2.8 and 7.1 [37].

- *Testing capacity* The number of confirmed cases is usually determined by testing, which is biased toward severe cases in some countries like France. In contrast, the testing is aimed at a larger group in some other countries implementing massive testing programs, such as South Korea and Iceland. Based on antibody tests performed on the general population, several reports show that the actual number of infected people is much larger than the reported value [38, 39].
- *Testing protocol* The testing protocols and accuracy may also have a large impact on the results. Depending on the testing protocols used, in some instances, false positive results have been obtained. In other words, someone without the disease tested positive, probably because they were infected with some other coronavirus. There have been several reports raising this issue [40]. On the other hand, false negative results may also exist and seem to be more prevalent than false positives.
- *Reporting system and time* Data also rely on the efficiency of the reporting system. Tests are conducted sequentially over time, and the reported data may be adjusted afterward. They do not represent a snapshot of a day in time. For instance, the Netherlands National Institute for Public Health and the Environment clearly states that some of the positive results are only reported one or a few days later and there might be corrections for the past data, so the numbers from a few days ago are sometimes adjusted [41].

Therefore, the real number of cases in the population is likely to be many multiples higher than those computed from confirmed tests, and the number of confirmed cases can only reflect the real situation of the outbreak to some degree. This may also partly explain that the Logistic model fails to capture the growth dynamics at the early stage in most provinces in China, as shown in Sect. 5, likely due to the potential underreporting at the beginning.

Moreover, factors like heterogeneous demographics, government response, people's fundamental health situations in different countries could lead to significantly different susceptibility to SARS-CoV-2. Numerous evidences are showing that the main vulnerable group in this COVID-19 pandemic is the older population or people with preexisting chronic diseases, such as cardiovascular disease, hypertension, diabetes and with obesity [42–45]. The discipline and distanciation culture in some Asian countries may also have contributed to the low number of confirmed cases there. All these factors affect the susceptibility of individuals and the transmission networks, which are not easy to handle in a microscopic model.

Therefore our top-down approach could be useful in quantifying and understanding the current status of the outbreak progress and making short-term predictions, while they tend to underestimate the final confirmed numbers in the long term, which has a higher uncertainty. Thus, the three Logistic type models could only provide a lower bound for the future scenarios. As more data become available, we anticipate a more accurate picture of the final numbers, as showed, for instance, by the converged ensemble distributions in Austria and Switzerland with small variance. As a last remark, even if the true number of cases is a multiple of the reported number of confirmed cases, as long as the multiple does not vary too much for a given country, our analysis remains pertinent to ascertain the outbreak progress and nature of the epidemic dynamics.

## 8 Conclusion

In this paper, we calibrated the logistic growth model, the generalized logistic growth model, the generalized growth model and the generalized Richards model to the reported number of confirmed cases in the Covid-19 epidemics from January 19 to March 10 for the whole of China and for the 29 provinces in China. This has allowed us to draw some lessons useful to interpret the results of a similar modeling exercise performed on other countries, which are at less advanced stages of their outbreaks. Our analysis dissects the development of the epidemics in China and the impact of the drastic control measures both at the aggregate level and within each province. We documented four phases: I—early stage outbreak (January 19–January 24, 6 days), II—fast growth phase approaching the peak of the incidence curve (January 25–February 1, 8 days), III—slow growth phase approaching the end of the outbreak (February 2–February 14, 13 days) and IV—the end of the outbreak (February 15–8 March). We quantified the initial reactions and ramping up of control measures on the dynamics of the epidemics and unearthed an inverse relationship between the number of days from peak to the quasi-end and the duration from start to the peak of the epidemic among the 29 analyzed Chinese provinces. We identified the signatures of the dynamic for the exemplary developments in Zhejiang and Henan provinces and the heterogeneity of the development of the epidemic and responses across various other provinces. We found a strong correlation between the initial and total confirmed numbers of infected cases and travel index quantifying the mobility between provinces.

For countries that are in the middle of their outbreak, we constructed three scenarios to make future projections. We demonstrated that outbreaks in 14 countries (mostly in West Europe) have ended, while resurgences of cases are identified in several countries. The estimated parameters from the GRM are compared with the Chinese data, showing that European countries had much slower after-peak trajectories, possibly due to the earlier and stricter containment measures in China. We found that 5 countries (Chile, Canada, Afghanistan, Qatar and Belarus) have entered into the final stage of the epidemic, while countries in the South Hemisphere and developing areas (Brazil, Argentina, Sweden, Mexico, Israel, India, Dominican Republic, Philippines and Iran) are still at an early stage of the outbreaks as of July 24, 2020. The USA, Israel, Philippines, Iran and several other countries are experiencing a significant second wave of outbreak, which is out of the scope of our models. We use ensemble distributions based on bootstrapping simulations to demonstrate the convergences and uncertainties of our scenario-based predictions. Our quantitative analysis allows estimating the progress of the outbreak in different countries, differentiating between those that are at a quite advanced stage and close to the end of the epidemics from those that are still in the middle of it. This is useful for scientific colleagues and decision makers to understand the different dynamics and status of countries and regions in a simple and transparent way.

**Acknowledgements** Open access funding provided by Swiss Federal Institute of Technology Zurich. We benefitted from many stimulating discussions and exchanges with Michael Schatz, Peter Cauwels, Dmitry Chernov, Euan Mearns, Pengcheng Li and Yixuan Zhang.

**Authors' contributions** KW and DS designed the research. KW and QW performed the data analysis. All authors wrote, read and approved the final manuscript.

**Funding** No funding information is applicable.

**Availability of data and material** The datasets generated and analyzed during the current study are available in the Github repository, https://github.com/kezida/covid-19-logistic-paper.

**Code availability** Code is available upon request.

**Compliance with ethical standards**

**Conflict of interest** The authors declare that they have no conflict of interest.

## References

1. Dye, C., Gay, N.: Modeling the SARS epidemic. Science **300**(5627), 1884–1885 (2003). https://doi.org/10.1126/science.1086925
2. Laguzet, L., Turinici, G.: Individual vaccination as Nash equilibrium in a SIR model with application to the 2009–2010 influenza A (H1N1) epidemic in France. Bull. Math. Biol. **77**(10), 1955–1984 (2015). https://doi.org/10.1007/s11538-015-0111-7

K. Wu et al.

3. Zhao, Z., Calderón, J., Xu, C., Zhao, G., Fenn, D., Sornette, D., Crane, R., Hui, P.M., Johnson, N.F.: Effect of social group dynamics on contagion. Phys. Rev. E **81**(5), 056107 (2010). https://doi.org/10.1103/PhysRevE.81.056107
4. Lekone, P.E., Finkenstädt, B.F.: Statistical inference in a stochastic epidemic SEIR model with control intervention: Ebola as a case study. Biometrics **62**(4), 1170–1177 (2006). https://doi.org/10.1111/j.1541-0420.2006.00609.x
5. Smieszek, T., Balmer, M., Hattendorf, J., Axhausen, K.W., Zinsstag, J., Scholz, R.W.: Reconstructing the 2003/2004 H3N2 influenza epidemic in Switzerland with a spatially explicit, individual-based model. BMC Infect. Dis. **11**(1), 115 (2011). https://doi.org/10.1186/1471-2334-11-115
6. Chang, S.L., Piraveenan, M., Pattison, P., Prokopenko, M.: Game theoretic modelling of infectious disease dynamics and intervention methods: a review. J. Biol. Dyn. **14**(1), 57–89 (2020). https://doi.org/10.1080/17513758.2020.1720322
7. Chang, S.L., Harding, N., Zachreson, C., Cliff, O.M., Prokopenko, M.: Modelling Transmission and Control of the COVID-19 Pandemic in Australia (2020). arXiv preprint arXiv:2003.10218
8. Prem, K., Liu, Y., Russell, T.W., Kucharski, A.J., Eggo, R.M., Davies, N., Flasche, S., Clifford, S., Pearson, C.A.B., Munday, J.D., Abbott, S., Gibbs, H., Rosello, A., Quilty, B.J., Jombart, T., Sun, F., Diamond, C., Gimma, A., van Zandvoort, K., Funk, S., Jarvis, C.I., Edmunds, W.J., Bosse, N.I., Hellewell, J., Jit, M., Klepac, P.: The effect of control strategies to reduce social mixing on outcomes of the COVID-19 epidemic in Wuhan, China: a modelling study. Lancet Publ. Health (2020). https://doi.org/10.1016/S2468-2667(20)30073-6
9. Hou, C., Chen, J., Zhou, Y., Hua, L., Yuan, J., He, S., Guo, Y., Zhang, S., Jia, Q., Zhao, C., Zhang, J., Xu, G., Jia, E.: The effectiveness of quarantine of Wuhan city against the Corona Virus Disease 2019 (COVID-19): a well-mixed SEIR model analysis. J. Med. Virol. (2020). https://doi.org/10.1002/jmv.25827
10. Boldog, P., Tekeli, T., Vizi, Z., Dénes, A., Bartha, F.A., Röst, G.: Risk assessment of novel coronavirus COVID-19 outbreaks outside China. J. Clin. Med. **9**(2), 571 (2020). https://doi.org/10.3390/jcm9020571
11. Rocklöv, J., Sjödin, H., Wilder-Smith, A.: COVID-19 outbreak on the Diamond Princess cruise ship: estimating the epidemic potential and effectiveness of public health countermeasures. J. Travel Med. (2020). https://doi.org/10.1093/jtm/taaa030
12. Ferguson, N., Laydon, D., Nedjati Gilani, G., Imai, N., Ainslie, K., Baguelin, M., Bhatia, S., Boonyasiri, A., Cucunuba Perez, Z., Cuomo-Dannenburg, G., Dighe, A., Dorigatti, I., Fu, H., Gaythorpe, K., Green, W., Hamlet, A., Hinsley, W., Okell, L.C., van Elsland, S., Thompson, H., Verity, R., Volz, E., Wang, H., Wang, Y., Walker, P.G., Walters, C., Winskill, P., Whittaker, C., Donnelly, C.A., Riley, S., Ghani., A.C.: Report 9: impact of non-pharmaceutical interventions (NPIs) to reduce COVID19 mortality and healthcare demand (2020). https://doi.org/10.25561/77482
13. Sornette, D.: Predictability of catastrophic events: material rupture, earthquakes, turbulence, financial crashes, and human birth. Proc. Natl. Acad. Sci. **99**(suppl 1), 2522–2529 (2002). https://doi.org/10.1073/pnas.022581999
14. Sornette, D.: Critical Phenomena in Natural Sciences: Chaos, Fractals, Selforganization and Disorder: Concepts and Tools. Springer, Berlin (2006)
15. Israeli, N., Goldenfeld, N.: Computational irreducibility and the predictability of complex physical systems. Phys. Rev. Lett. **92**(7), 074105 (2004). https://doi.org/10.1103/PhysRevLett.92.074105
16. Gourieroux, C., Jasiak, J.: Time Varying Markov Process with Partially Observed Aggregate Data; An Application to Coronavirus (2020). arXiv preprint arXiv:2005.04500
17. Ekum, M., Ogunsanya, A.: Application of hierarchical polynomial regression models to predict transmission of COVID-19 at global level. Int. J. Clin. Biostat. Biom. **6**, 027 (2020)
18. Chowell, G., Hincapie-Palacio, D., Ospina, J., Pell, B., Tariq, A., Dahal, S., Moghadas, S., Smirnova, A., Simonsen, L., Viboud, C.: Using phenomenological models to characterize transmissibility and forecast patterns and final Burden of Zika epidemics. PLoS Curr. (2016). https://doi.org/10.1371/currents.outbreaks.f14b2217c902f453d9320a43a35b9583
19. Chowell, G.: Fitting dynamic models to epidemic outbreaks with quantified uncertainty: a primer for parameter uncertainty, identifiability, and forecasts. Infect. Dis. Model. **2**(3), 379–398 (2017). https://doi.org/10.1016/j.idm.2017.08.001
20. Viboud, C., Simonsen, L., Chowell, G.: A generalized-growth model to characterize the early ascending phase of infectious disease outbreaks. Epidemics **15**, 27–37 (2016). https://doi.org/10.1016/j.epidem.2016.01.002
21. Chowell, G., Tariq, A., Hyman, J.M.: A novel sub-epidemic modeling framework for short-term forecasting epidemic waves. BMC Med. **17**(1), 1–18 (2019). https://doi.org/10.1186/s12916-019-1406-6
22. Chowell, G., Luo, R., Sun, K., Roosa, K., Tariq, A., Viboud, C.: Real-time forecasting of epidemic trajectories using computational dynamic ensembles. Epidemics. **30**, 100379 (2020). https://doi.org/10.1016/j.epidem.2019.100379
23. Beaubien, J.: China Enters The Next Phase of Its COVID-19 Outbreak: Suppression (2020). https://www.npr.org/sections/goatsandsoda/2020/04/03/826140766/china-enters-the-next-phase-of-its-covid-19-outbreak-suppression. Accessed 24 Apr 2020
24. European Centre for Disease Prevention and Control (ECDC): Situation update worldwide (2020). https://www.ecdc.europa.eu/en/geographical-distribution-2019-ncov-cases. Accessed 24 Apr 2020
25. Remuzzi, A., Remuzzi, G.: COVID-19 and Italy: what next? Lancet **395**(10231), 1225–1228 (2020). https://doi.org/10.1016/S0140-6736(20)30627-9
26. Ma, J., Dushoff, J., Bolker, B.M., Earn, D.J.: Estimating initial epidemic growth rates. Bull. Math. Biol. **76**(1), 245–260 (2014). https://doi.org/10.1007/s11538-013-9918-2
27. Kermack, W.O., McKendrick, A.G.: A contribution to the mathematical theory of epidemics. Proc. R. Soc. Lond. Ser. A Contain. Pap. Math. Phys. Character **115**(772), 700–721 (1927). https://doi.org/10.1098/rspa.1927.0118

28. Richards, F.J.: A flexible growth function for empirical use. J. Exp. Bot. **10**(2), 290–301 (1959). https://doi.org/10.1093/jxb/10.2.290
29. Neher, D.A., Campbell, C.L.: Underestimation of disease progress rates with the logistic, monomolecular, and gompertz models when maximum disease intensity is less than 100 percent. Phytopathology **82**(8), 811–814 (1992)
30. Pell, B., Kuang, Y., Viboud, C., Chowell, G.: Using phenomenological models for forecasting the 2015 Ebola challenge. Epidemics **22**, 62–70 (2018)
31. Tian, Y.: The Tough Time Through the Chinese New Year (in Chinese: 既过年关, 也过难关) (2020). https://web.archive.org/web/20200125183422/http://www.xinhuanet.com/politics/2020-01/25/c_1125501347.htm. Accessed 25 Jan 2020
32. He, X.: How Strong is Henan in Preventing and Controlling COVID-19? (2020) (in Chinese: 防控肺炎病毒,"硬核"河南究竟有多硬核?). http://www.nbd.com.cn/articles/2020-01-25/1402907.html. Accessed 25 Jan 2020
33. Lai, S., Bogoch, I.I., Watts, A., Khan, K., Li, Z., Tatem, A.: Preliminary Risk Analysis of 2019 Novel Coronavirus Spread Within and Beyond China (2020). https://www.pentapostagma.gr/sites/default/files/2020-02/worldpop-coronavirus-spread-risk-analysis-v1-25jan.pdf. Accessed 25 Feb 2020
34. Ying, S., Li, F., Geng, X., Li, Z., Du, X., Chen, H., Chen, S., Zhang, M., Shao, Z., Wu, Y., Syeda, M.Z., Yan, F., Che, L., Zhang, B., Lou, J., Wang, S., Chen, Z., Li, W., Shen, Y., Chen, Z., Shen, H.: Spread and Control of COVID-19 in China and Their Associations with Population Movement, Public Health Emergency Measures, and Medical Resources. medRxiv, 2020.2002.2024.20027623 (2020). https://doi.org/10.1101/2020.02.24.20027623
35. Jarlov, H.: Anti-SARS-CoV-2 Screening (2020). https://docs.google.com/spreadsheets/d/17Tf1Ln9VuE5ovpnhLRBJH-33L5KRaiB3NhvaiF3hWC0/edit#gid=0. Accessed 18 Jul 2020
36. Chair of Entrepreneurial Risks, E.Z.: COVID-19 Daily Report (2020). https://er.ethz.ch/Covid-19/Dailyforecasts.html
37. Tsang, T.K., Wu, P., Lin, Y., Lau, E.H., Leung, G.M., Cowling, B.J.: Effect of changing case definitions for COVID-19 on the epidemic curve and transmission parameters in mainland China: a modelling study. Lancet Publ. Health (2020). https://doi.org/10.1016/S2468-2667(20)30089-X
38. Bendavid, E., Mulaney, B., Sood, N., Shah, S., Ling, E., Bromley-Dulfano, R., Lai, C., Weissberg, Z., Saavedra, R., Tedrow, J., Tversky, D., Bogan, A., Kupiec, T., Eichner, D., Gupta, R., Ioannidis, J., Bhattacharya, J.: COVID-19 Antibody Seroprevalence in Santa Clara County, California. medRxiv, 2020.2004.2014.20062463 (2020). https://doi.org/10.1101/2020.04.14.20062463
39. Streeck, H., Hartmann, G., Exner, M., Schmid, M.: Preliminary Results and Conclusions of the COVID-19 Case Cluster Study (Gangelt municipality) (in Germany: Vorläufiges Ergebnis und Schlussfolgerungen der COVID-19 Case-ClusterStudy (Gemeinde Gangelt)) (2020). https://www.land.nrw/sites/default/files/asset/document/zwischenergebnis_covid19_case_study_gangelt_0.pdf. Accessed 23 Apr 2020
40. Deeks JJ, D.J., Takwoingi Y, Davenport C, Spijker R, Taylor-Phillips S, Adriano A, Beese S, Dretzke J, Ferrante di Ruffano L, Harris IM, Price MJ, Dittrich S, Emperador D, Hooft L, Leeflang MMG, Van den Bruel A: Antibody tests for identification of current and past infection with SARS-CoV-2. Cochrane Database Syst. Rev. **2020**(6), Art. No.: CD013652 (2020). https://doi.org/10.1002/14651858.cd013652
41. Environment, T.N.N.I.f.P.H.a.t.: Development of COVID-19 in Graphs (2020). https://www.rivm.nl/coronavirus-covid-19/grafieken. Accessed 23 Apr 2020
42. Zhou, F., Yu, T., Du, R., Fan, G., Liu, Y., Liu, Z., Xiang, J., Wang, Y., Song, B., Gu, X., Guan, L., Wei, Y., Li, H., Wu, X., Xu, J., Tu, S., Zhang, Y., Chen, H., Cao, B.: Clinical course and risk factors for mortality of adult inpatients with COVID-19 in Wuhan, China: a retrospective cohort study. Lancet **395**(10229), 1054–1062 (2020). https://doi.org/10.1016/S0140-6736(20)30566-3
43. Richardson, S., Hirsch, J.S., Narasimhan, M., Crawford, J.M., McGinn, T., Davidson, K.W., Consortium, a.t.N.C.-R.: Presenting Characteristics, Comorbidities, and Outcomes Among 5700 Patients Hospitalized With COVID-19 in the New York City Area. JAMA (2020). https://doi.org/10.1001/jama.2020.6775
44. Wang, D., Hu, B., Hu, C., Zhu, F., Liu, X., Zhang, J., Wang, B., Xiang, H., Cheng, Z., Xiong, Y., Zhao, Y., Li, Y., Wang, X., Peng, Z.: Clinical Characteristics of 138 Hospitalized Patients With 2019 Novel Coronavirus-Infected Pneumonia in Wuhan, China. JAMA **323**(11), 1061–1069 (2020). https://doi.org/10.1001/jama.2020.1585
45. Guan, W.-J., Ni, Z.-Y., Hu, Y., Liang, W.-H., Ou, C.-Q., He, J.-X., Liu, L., Shan, H., Lei, C.-L., Hui, D.S.C., Du, B., Li, L.-J., Zeng, G., Yuen, K.-Y., Chen, R.-C., Tang, C.-L., Wang, T., Chen, P.-Y., Xiang, J., Li, S.-Y., Wang, J.-L., Liang, Z.-J., Peng, Y.-X., Wei, L., Liu, Y., Hu, Y.-H., Peng, P., Wang, J.-M., Liu, J.-Y., Chen, Z., Li, G., Zheng, Z.-J., Qiu, S.-Q., Luo, J., Ye, C.-J., Zhu, S.-Y., Zhong, N.-S.: Clinical characteristics of coronavirus disease 2019 in China. N. Engl. J. Med. **382**(18), 1708–1720 (2020). https://doi.org/10.1056/nejmoa2002032